\documentclass[prl,twocolumn]{revtex4}

\usepackage{graphicx}
\usepackage{amsmath}
\usepackage{dcolumn}  
\usepackage{bm}       
\usepackage{color}
\usepackage{epstopdf}
\usepackage{amssymb}
\usepackage{epsfig,times}
\usepackage{amstext}
\usepackage{latexsym}
\usepackage{hyperref}
\usepackage{txfonts}

\begin{document}

\title{Orthogonality catastrophe as a consequence of probe embedding in an ultra-cold Fermi gas}

\author{J.~Goold$^{1,2}$, T.~Fogarty$^2$, N.~Lo~Gullo$^2$, M.~Paternostro$^3$, and Th.~Busch$^2$}

\affiliation{ $^1$Clarendon Laboratory, University of Oxford, United Kingdom,\\
  $^2$Physics Department, University College Cork, Cork, Ireland,\\
  $^3$Centre for Theoretical Atomic, Molecular and Optical Physics,
  Queen's University Belfast, Belfast BT7 1NN, UK }

\date{\today}

\begin{abstract}
  We investigate the behavior of a two-level atom
  coupled to a one dimensional, ultra-cold Fermi gas. The {\it sudden} switching
  on of the impurity-gas scattering leads to the loss of any
  coherence in the initial state of the impurity and we show that the exact
  dynamics of this process is strongly influenced by the effect of the
  {\it orthogonality catastrophe} within the gas. We highlight the
  relationship between the Loschmidt echo and the retarded Green's
  function - typically used to formulate the dynamical theory of the
  catastrophe - and demonstrate that the effect is reflected in the impurity
  dynamics. The expected broadening of the spectral function can be observed 
  using Ramsey interferometry on the two level atom.
\end{abstract}
\maketitle 

In the past decade ultra-cold quantum gases have emerged as ideal
candidates for designing controllable experiments to simulate effects
in condensed matter physics \cite{Bloch:08}.  This success is likely
to receive another boost due to the recent emergence of a fundamental
new class of hybrid experimental systems. In these, two separate,
ultra-cold atomic systems are combined in such a way that their
coupling can be externally controlled and their states independently
measured, therefore allowing detailed investigations into the theory
of quantum interactions and decoherence. Existing examples of such
systems are single spin impurities embedded in ultra-cold Fermi
gases~\cite{Zwierlein:09,Salamon:09} and the combination of
neutral  \cite{Widera:10,Bloch:10} or charged single atoms
\cite{kohlion:10,Denschlag:2010} with Bose-Einstein
condensates. These endeavors offer the possibility for controlled
simulation of many different system-environment models synonymous with
condensed matter physics and non-equilibrium statistical physics
\cite{mahan:00}.

Here we show that a fundamental and well known quantum many-body
effect, the Anderson orthogonality catastrophe (OC), can play an
important role in ultra-cold, coupled systems \cite{Anderson:67}. We
consider a single two-level system (impurity) embedded into a harmonically
trapped, ultra-cold Fermi gas and demonstrate how the overlap
between the many-body wavefunctions of the Fermi sea before and after
a transition in the impurity vanishes as the number of particle
 and/or the scattering strength is increased. 
This signals the onset of OC and we show that it can be
observed by looking at the dynamical dephasing features of the impurity
alone.  Our study identifies the OC as a significant effect even in
mesoscopic systems and it represents a prime example of how properties 
of an out-of-equilibrium systems with many degrees of freedom can be 
inferred by looking at a simpler auxiliary system.

Let us briefly revisit the original idea of Anderson
\cite{Anderson:67} by considering the ground state of a
non-interacting, spin-polarized Fermi gas in a hard-wall,
spherically-symmetric box at zero temperature. The many-particle
wave-function of the gas is given by the Slater determinant of the
radial single-particle eigenstates $\psi_n(k_j,x_j)$ as
\begin{equation}
  \Psi(r_1,r_2,\dots,r_N) =\frac{1}{\sqrt N!} 
                            \det_{(n,j)=(0,1)}^{(N-1,N)}\psi_n(k_j,r_j)\;,
\label{eq:psii}
\end{equation}
where $r_j$ ($k_j$) is the coordinate (wavenumber) of the
$j^{\text{th}}$ particle. For spherical symmetry ($l=0$) the
eigenstates are given by the Bessel functions
$\psi_n(k_j,r_j)=\frac{\sin(k_j r_j)}{k_j r_j}$.  Consider now the
same system, but in the presence of a static perturbation.
Intuitively, the single-particle states are deformed and if the
perturbation is highly localised, the new states can be written
asymptotically as $\psi^{'}_n(k_j,r_j){\sim}\frac{\sin(k_j
  r_j+\delta)}{k_j r_j}(1-\frac{r_j}{R})$, where $\delta$ is an s-wave
phase shift and $R$ is the radius of the spherical box. This, in turn,
leads to a modified Fermi sea $\Psi^{'}$ and the overlap between the
perturbed and unperturbed many-body states is given by
$\nu{=}\langle\Psi^{'}|\Psi\rangle{=}\det [A_{nm}]$, where
$A_{nm}=\int{\psi_{n}(r)\psi^{'}_m(r)}\;dr$. When evaluating the
overlap integral, one finds $\nu{\propto}N^{-\frac{\alpha}{2}}$, where
$\alpha=\frac{2\delta^2}{\pi^2}$ goes to zero for $N$ and/or $\delta$
sufficiently large~\cite{Anderson:67}.

While Anderson's original work involved stationary states, the
creation of a perturbed many-body state is, in general, a
time-dependent process. The dynamical theory of the OC was developed
by Nozi\`eres and De Dominicis \cite{Nozieres:69}, who subsequently calculated the
transient response of a Fermi sea after the {\it sudden} switching on
of a core-hole in a metal. A direct manifestation of the OC can then
be observed in the single-particle spectrum of the Fermi gas
\begin{equation}
  \label{eq:spectrum}
  A(\omega){=}2\Re\int^{\infty}_{-\infty}\!\!\!dt\;e^{i(\omega-\omega_T)t}\nu(t)\;,
\end{equation} 
where, $\nu(t)=\langle \Psi|e^{i\hat Ht}e^{-i\hat
  H^{'}t}|\Psi\rangle$ is the propagator of the core hole's retarded
Green's function at zero temperature
\begin{eqnarray}
  \label{eq:greens1}
  G(t)&=&-ie^{-it\omega_T}\Theta(t)
          \langle \Psi|e^{i\hat Ht}e^{-i\hat H^{'}t}|\Psi\rangle. 
\end{eqnarray} 
Here $\Theta(t)$ is the Heaviside step function and $\Psi$ is the
initial equilibrium state of the Fermi system, governed by the
Hamiltonian $\hat H$. The threshold frequency for the creation of the
hole in the valence band of the metal is given by $\omega_T$ and the
subsequent evolution of the Fermi sea in the presence of the impurity
is given by $\hat H^{'}$ \cite{mahan:00}. In the absence of the hole,
the single-particle spectrum of the homogeneous non-interacting Fermi
gas is a Dirac-$\delta$ function peaked at the Fermi
energy. Experimentally, it was shown using X-ray spectroscopy that the
presence of the OC broadens this spectrum, thus signaling a dramatic
change in the fundamental excitations of the system.

The evaluation of the Green's function in Eq. ~\eqref{eq:greens1}
amounts to calculating the single particle state overlaps if one can 
solve the single particle problem for both the perturbed and the unperturbed system.
In this case in fact the evolved many-body wavefunctions are simply 
given by combinations of Slater determinants of the evolved single 
particles states $\psi_n(x,t)$ and $\psi_n'(x,t)$.
The overlap $\nu(t)$ between the two many-body wavefunctions is thus:
\begin{equation}
  \label{eq:greens}
  \nu(t)=\langle \Psi|e^{i\hat Ht}e^{-i\hat H^{'}t}|\Psi\rangle =\det[A_{nm}(t)]
\end{equation} 
with $A_{nm}(t){=}\int\psi^{'}_n(x,t)\psi_m(x,t)dx$.

In what follows we demonstrate how the physics of the OC influences
the dynamics of a single auxiliary two-level system which is coupled
to the environment of a non-interacting Fermi gas in a harmonic trap. As our system we
choose a highly localised, neutral atom~\cite{Alessio:05,Bruderer:06},
whose relevant two levels, $|g\rangle$ and $|e\rangle$, are assumed to
be separated by the energy $\hbar\Omega$, so that the free Hamiltonian
reads $H_{s}{=}\frac{\hbar\Omega}{2}(|e\rangle\langle
e|-|g\rangle\langle g|)$. The environment is then described by
\begin{equation}
  \label{eq:Hamiltonian_fermions}
  \hat{H}=\int\hat\Psi^\dagger(x)\left(-\frac{\hbar^2}{2m}\frac{d^2}{dx^2}
           +\frac{1}{2}m\omega^2 x^2\right)\hat\Psi(x)dx,
\end{equation} 
where $\hat\Psi^\dagger(x)$ is the fermionic field creation
operator. Note that  we are choosing a
one-dimensional system here which is sufficent to demonstrate the fundemental effects. At sufficiently low temperatures s-wave
scattering is the dominating interaction process between the Fermi gas
and the atom and for simplicity (without affecting the generality of
our discussion) we assume that only $|e\rangle$ has a finite
(positive) s-wave scattering length, while $|g\rangle$ does not
interact with the environment. 
Assuming that a confining potential 
strongly localizes the impurity's wave-function so that the kinetic
energy can be neglected then leads to the following interaction
Hamiltonian for the gas
\begin{equation}
  \label{eq:Hamiltonian_interaction}
  \hat{H}_I=\kappa\int\hat\Psi^\dagger(x)\delta(x-d)
                                         \hat\Psi(x)dx,
\end{equation}
where we have used the standard pseudo-potential approximation for the
scattering interaction: the scattering potential only acts at position
$d$ in the gas, with a strength $\kappa$ that can be related to the
previously mentioned scattering phase-shift $\delta$ \cite{Landau}. In
this work we will specifically focus on the case $d=0$.
The analogy with the situation typically
considered in the context of Anderson's OC theory should now be
apparent: the localised spatial interaction of the impurity with the
ultra-cold gas plays a role analogous to the interaction of the core
hole with the rest of the electrons in a metal. A key point to stress
is that here, in contrast to the case of a metal, we have a far
smaller number of particles in the environment, which could in
principle compromise the observability of the OC effects. However,
ultra-cold atomic systems allow for the possibility to tune the s-wave
scattering length to an arbitrarily large value by means of a
Feshbach resonance, thus compensating for the lack of
particles participating in such dynamics. We will show that this
offers the possibility of observing the OC effect even in mesoscopic
systems.

Let us start by assuming that, at time $t<0$, the atom is prepared in
the state $|g\rangle$ and the Fermi gas is in its ground state
$|\Psi\rangle$. As $|g\rangle$ does not interact with the Fermi gas,
the collective state of the hybrid system can be written as
$|\Phi\rangle=|g\rangle\otimes|\Psi\rangle$. At $t=0$, a properly set interaction between the 
atom and a classical laser field prepares the former in $(|g\rangle+|e\rangle)/\sqrt2$ and the perturbed Fermi sea evolves according to $\hat
H_I$, driving the overall system into a correlated state of the form
\begin{align}
  \label{eq:entangledqf}
  |\Phi^{'}\rangle=&\left(|g\rangle\otimes e^{-i\hat H
      t}|\Psi\rangle+|e\rangle\otimes e^{-i(\hat H+\hat
      H_I)t}|\Psi\rangle\right)/\sqrt 2.
\end{align}
The state of the environment now depends on atomic
state with $|\Psi^{'}_g(t)\rangle=e^{-i\hat Ht}|\Psi\rangle$ for the
non-interacting part and $|\Psi^{'}_e(t)\rangle=e^{-i(\hat H+\hat
  H_I)t}|\Psi\rangle$ in the presence of scattering. The time-dependent density
matrix of the impurity $\rho_s(t)$ can straightforwardly be evaluated by
tracing out the environment. One immediately finds that the
coherences of the reduced state are proportional to the scalar product
\begin{equation}
  \label{eq:nut}
  \langle\Psi^{'}_{g}(t)|\Psi^{'}_e(t)\rangle=\langle \Psi|e^{it\hat H}
         e^{-it(\hat H+\hat H_I)}|\Psi\rangle=\nu(t).
\end{equation}
The equivalence between this expression and the time-propagator
$\nu(t)$ in Eq.~(\ref{eq:greens}) is immediately evident, thus providing a direct link
between the decoherence of an atom in a fermionic environment and the
phenomenon of Anderson OC. Furthermore,
$|\nu(t)|^2{\equiv}|\langle\Psi^{'}_g(t)|\Psi^{'}_e(t)\rangle|^2$ is the
so-called Loschmidt echo $L(t)$~\cite{Zanardi:06,le}, which can be used
as a tool for the quantitative study of decoherence processes due to
many-body environments~\cite{zurek:03, Quan}. We shall go back to the study of the echo 
later.  

The above argument holds for situations in which the Fermi gas is initially prepared in 
a pure state. However it is often the case that the gas 
has some thermal component and its quantum state is mixed. In this case, the state  of the environment at 
$t=0$ is described by a density matrix and we find the overlap to be 
 \begin{equation}
 \label{eq:temp}
\nu(t)={\rm Tr}[\hat{U}^{'}(t)\hat{\rho}\hat{U}(-t)]=\sum_{n,m,l}C_{l}\Lambda_{m,n}^{*}\Lambda_{m,l}\; e^{-i\Delta_{m,n}t}
\end{equation}
with $\Delta_{m,n}{=}E^{'}_{m}{-}E_{n}$ the difference between the many-body energies $E_n$ and $E'_m$,  $C_{l}=e^{E_{l}/k_{B}T}/Z$ (here $Z$ is the partition function) and $\Lambda_{m,n}$ the overlap between the excited many-body states of the system (see Ref.~\cite{epaps} for further details). In the following, for the sake of clarity, we present all results in natural units and thus scale energies in terms of $\hbar\omega$, lengths in terms of the harmonic
trap length $l_{0}{=}\sqrt{\hbar/m\omega}$, and time in units of
the inverse trapping frequency $\omega^{-1}$. From this it follows
that $\kappa$ is scaled in units of $ l_{0}/\hbar\omega$. 

Given the formal connection between $\nu(t)$ and the impurity's dynamics, we can
quantify the degree of quantum correlations embodied by entanglement
within the state in Eq.~\eqref{eq:entangledqf} by means of the von
Neumann entropy $S(t){=-}\sum_i\lambda_i(t)\log_2\lambda_i(t)$, where
$\lambda_{i}(t)$ are the time-dependent eigenvalues of $\rho_s(t)$.
\begin{figure}[tb]
  \includegraphics[width=\linewidth]{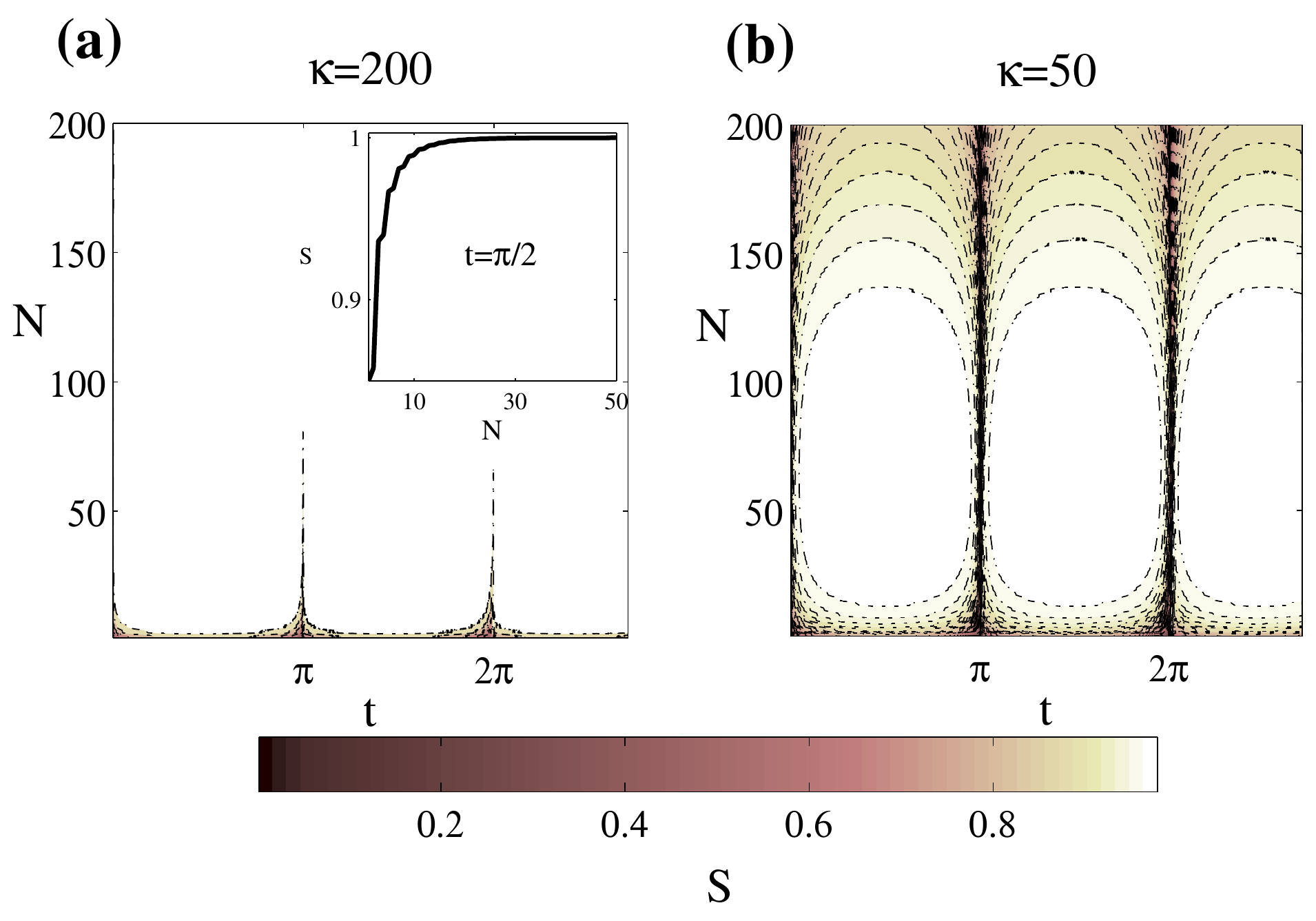}
  \caption{{\bf (a)} Time-dependent von Neumann entropy as a function of
   the particle number for $\kappa=200$. The inset shows a time slice at
    $t=\pi/2$. {\bf (b)} Time-dependent von Neumann entropy as a function
    of the particle number for $\kappa=50$.}
  \label{fig:vonneumann}
\end{figure}
The time dependent von Neumann entropy is shown in
Fig.~\ref{fig:vonneumann} for systems with different particle number
and for two different values of the interaction strengths. If the
interaction energy is at or above the Fermi energy, as is the case in 
Fig.~\ref{fig:vonneumann}{\bf (a)}, it can be seen that  after
the interaction is switched on, the coupled system evolves into a
fully entangled state (having $S{=}1$). This indicates that the many-particle
state, created after the disturbance is switched on, almost immediately
becomes orthogonal to the initial equilibrium state, as demanded by the
catastrophe effect. It is remarkable to note that, already for a small number of 
particles, the state of the atomic gas is not separable at any time following the
quench.
The inset shows the entropy at a fixed moment in time,
clearly indicating that the orthogonal state is already reached for a
mesoscopic particle numbers ($N\approx 15$ in the figure). An interesting point
to make here is that, provided  one has the ability to tune the
coupling to a large value, the qualitative features remain similar for
even smaller Fermi environments. This is in contrast to the case of a
metal where large particle numbers but only relatively weak
scattering are in order.  Fig.~\ref{fig:vonneumann}{\bf (b)} shows the von Neumann entropy
 for a weaker coupling of $\kappa=50$. In this case, 
full orthogonality only appears for larger particle numbers and a
maximally entangled state is achieved around the resonance of
$N=50$.

\begin{figure}[tb]
  \includegraphics[width=\linewidth]{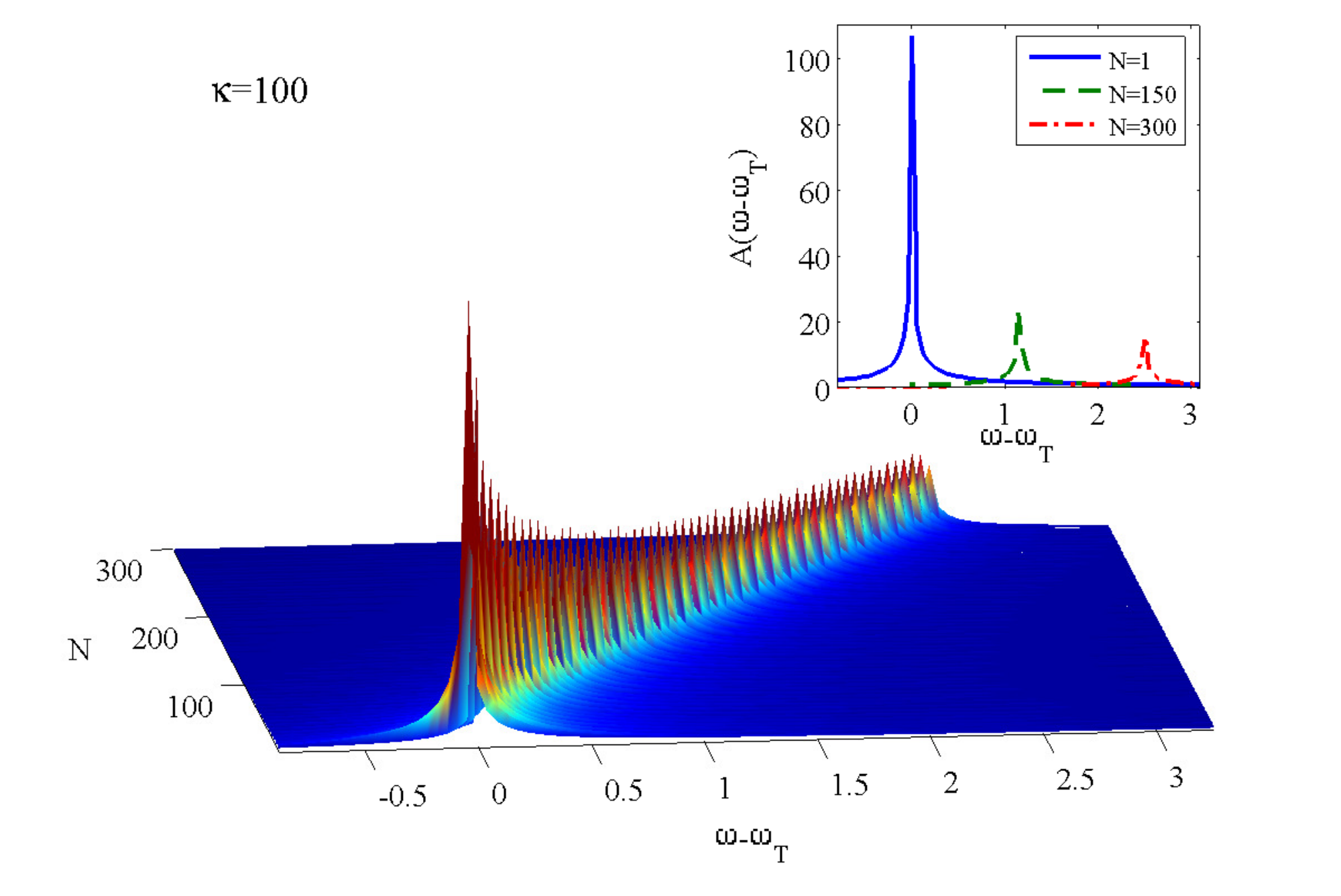}
  \caption{The spectral function $A(\omega)$ as a function of the particle number $N$
    for $\kappa=100$. The effect of OC is characterised by a flattening and a shift of 
the initally sharp and centralised peak, the area for all spectral functions is constant.}
  \label{fig:spectrum}
\end{figure}

Let us now show how the properties of our complex
system-environment state can be directly inferred from looking at the
system state only~\cite{mauro}. In particular, we suggest to use
Ramsey interferometry on the atom to measure the time dependent overlap, $\nu(t)$, and
from it the single-particle spectrum of the Fermi gas. As discussed
previously, this spectrum is known to be strongly affected by the OC
\cite{mahan:00}. Spectral information will therefore provide a
definite signature of OC which can be easily compared to the original
experiments in metals. 
Our scheme is based on a protocol put forward
in Ref.~\cite{dechiara:08}: after the creation of the entangled
atom-environment state, we allow the hybrid system to freely evolve
for a time $t$. During this time, a phase-shift gate is applied to the
atom, such that $|g\rangle \rightarrow |g\rangle$ and $|e\rangle
\rightarrow e^{i\phi}|e\rangle$, giving the state of the overall
system as $|\Psi(t)\rangle{=}(|g\rangle\otimes
e^{-i\hat Ht}|\Psi\rangle +{e^{i\phi}}|e\rangle\otimes
e^{-i(\hat H+H_I)t}|\Psi\rangle)/\sqrt2$. Using again a classical field, the state 
of the atom can be changed as $|g\rangle{\rightarrow}(|g\rangle{+}|e\rangle)/\sqrt2$, $|e\rangle{\rightarrow}(|g\rangle{-}|e\rangle)/\sqrt2$,
and we finally measure the probability for
the atom to be found in $|g\rangle$, which reads
\begin{equation}
  \label{eq:Ramsey}
  P_{g}(t,\phi)=[1+\cos(\phi)\nu_R(t)-\sin(\phi)\nu_I(t)]/2
\end{equation}
where $\nu(t)=\nu_R(t)+i\nu_I(t)$ is the overlap entering the OC
theory in Eq.~\eqref{eq:nut}. The overlap function $\nu(t)$ can be
extracted as a fitting parameter of Eq.~(\ref{eq:Ramsey}), which would be 
used as the best-fit function for the experimentally reconstructed $P_g(t,\phi)$. The latter 
would be achieved by measuring $P_g(t,\phi)$ (for instance through fluorescence resonance techniques)
at various value of the phase $\phi$. The single-particle spectrum $A(\omega)$ can then be obtained from the
Fourier transform on the time-dependent overlap $\nu(t)$, according to  Eq.~\eqref{eq:spectrum}. We show a typical spectrum in
Fig.~\ref{fig:spectrum} as a function of the particle number for $\kappa{=}100$ and within a frequency-window centered in $\omega_T$. 
One observes that $A(\omega)$ exhibits almost identical features to those observed via X-ray
absorption of metals~\cite{mahan:00}: as the number of particles in the gas increases, the peak frequency of the spectrum is renormalized (it is shifted towards the blue with respect to $\omega_T$) and the width of each peak broadens. As the area underneath each peak is constant, such broadening is consistent with a flattening effect.

Let us briefly discuss other experimental effects that can lead to a broadening of the spectrum. As one can see from Eq.~\eqref{eq:temp} the effect of a non-zero temperature of the gas is to introduce additional frequencies in the 
spectrum and thus to broaden it. Nonetheless, the latter is not the result of the OC but rather to the presence of initial
single particle excitations in the Fermi gas. Indeed the blue shift in the spectrum is also absent.
Another possible mechanism responsible for broadening is the finite size of the 
impurity. Such a source of imperfections can be included in our calculations by replacing the $\delta$-like interaction in Eq.~\eqref{eq:Hamiltonian_interaction} with a Gaussian potential with a characteristic width $\sigma$ while the $\delta$-like interaction only affects the initial even wavefunctions of the system,  leaving the odd ones unchanged \cite{buschhuyet}. A finite-size potential, on the other hand, would couple more states, thus broadening the spectrum.
Even though both the temperature and the finiteness of the impurity
produce a similar effects, it should be stressed that the latter is ``inherently part of
the OC.

\begin{figure}[tb]
  \includegraphics[width=\linewidth]{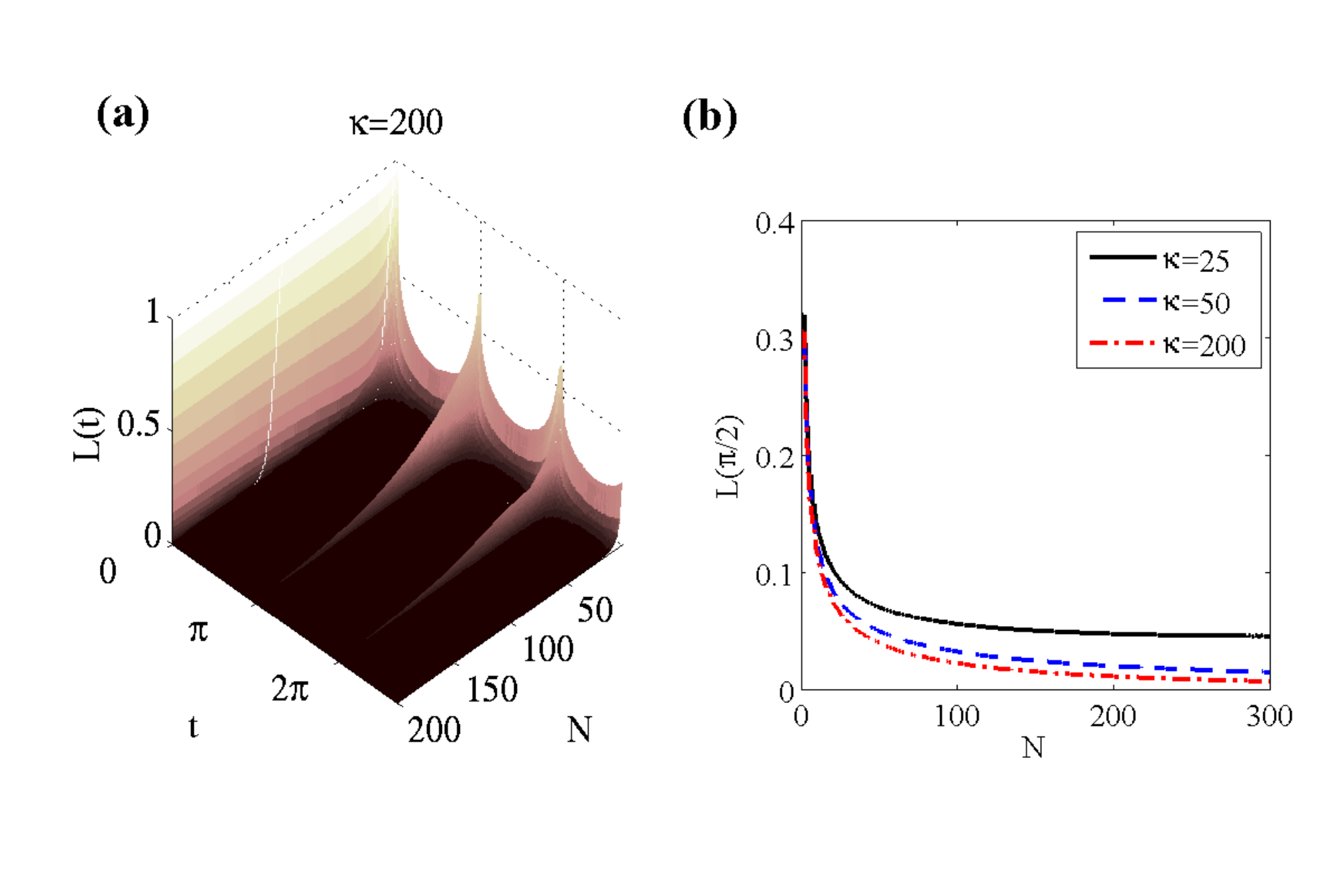}
  \caption{{\bf (a)} Loschmidt echo $L(t)$ as a function of the particle number
    $N$ for $\kappa=200$. {\bf (b)} Behaviour of $L(t)$ at time $t{=}{\pi}/{2}$ for
    $\kappa{=}25,50,100$.}
  \label{fig:Loschmidt}
\end{figure}

As remarked above, the OC manifests itself in an intriguing way in the
Loschmidt echo $L(t)$. The echo describes the environmental sensitivity to a generic
perturbation The echo corresponding to Fermi gases of various $N$ and $\kappa{=}200$ is shown in Fig.~\ref{fig:Loschmidt}.
As expected from our previous considerations, it decreases rapidly once the system size is above a moderate number. Fig.~\ref{fig:Loschmidt}{\bf (b)} 
shows the long-time behavior of $L(t)$ against $N$, highlighting its decreasing trend as the OC conditions are approached.
The revival in the echo are located at the time corresponding to the inverse of the particle-hole like resonances $E_n^{'}-E_0^{'}$ in the  (perturbed) Fermi gas (see Ref.~\cite{epaps}).
By measuring the dynamical overlap $\nu(t)$ one can thus probe the
single particle excitations of the system.

\noindent
{Let us finally comment on a realistic experimental implementation of our model.
A recent experiment utilised a species-selective dipole potential 
to trap (trap frequency 1kHz) and tightly localize an individual  $^{40}$K impurity 
in a one-dimensional gas of $^{87}$Rb (trap frequency 50Hz)~\cite{ssdp}. The average 
filling factor of the gas is $180$ atoms. Although
the atoms used in this specific example are bosonic, one can use a confinement-induced resonance to drive the
atoms into the fermionized Tonks-Girardeau regime, where the Loschmidt echo has recently been
shown to be equivalent to our result for non-interacting fermions ~\cite{croations}. 
Furthermore, when both species are prepared in the hyperfine state $|F{=}1,m_{f}{=}1\rangle$, one has the ability to 
adjust the interspecies scattering length almost at will utilising a
magnetic Feschbach resonance.
With the addition of suitable 
microwave pulses, the set-up in~\cite{ssdp} contains all of the ingredients for a direct 
experimental verification of our proposal.}

\noindent
In conclusion we have shown that the OC plays an important role in the dynamics of
coupled systems consisting of an ultra-cold atomic gas interacting
with a single two-level system. In this respect, we have quantitatively linked the OC to the
mechanism of decoherence undergone by the two-level system and signaled by the Loschmidt echo. It should be stressed that, beside pointing out the
exciting possibility to explore the OC in a realistic set-up, close to experimental state-of-the-art, and radically different from the one originally envisaged by Anderson, the scenario
addressed here allows us to demonstrate that, by manipulating a single
impurity, one can obtain highly nontrivial information about the
environmental behavior. Such information is invaluable for tasks of
environmental characterization and interaction-identification, thus
suggesting an ideal probe for testing ultracold atomic gases.  
In this sense, our proposal stands as the ultra-cold counterpart of the
hallmark experiments in the X-ray absorption spectrum of metals while
demonstrating, at the same time, the appropriateness of auxiliary quantum systems
as probes for ultra-cold quantum gases.

\begin{acknowledgments}
  We thank L.~Heaney, C.~Cormick and V.~Vedral for helpful
  discussions.  JG would like to acknowledge funding from an IRCSET
  Marie Curie International Mobility fellowship. This work was also
  supported by SFI under grant numbers 05/IN/I852, 05/IN/I852 NS and
 10/IN.1/I2979, IRCSET through the Embark Initiative RS/2009/1082 and  EPSRC (EP/G004579/1).
\end{acknowledgments}


\begin{thebibliography}{99}
\bibitem{le} In order to fully appreciate the utility of the Loschmidt
  echo in this context, it is instructive to calculate the purity of
  the system, $P(t)=\text{Tr}_s(\rho_s^{2})$, which gives
  $P(t)=1-\frac{1}{2}(1-L(t))$. When $L(t)\rightarrow 0$ then
  $P\rightarrow\frac{1}{2}$, thus indicating that the two-level system is
  maximally entangled with the environment. 
\end{thebibliography}
\end{document}